\documentstyle[12pt,aaspp4]{article}

\begin{document}

\lefthead{HAMAGUCHI}
\righthead{X-ray Flare from a Herbig Be star, MWC297}

\title{Large X-ray Flare from a Herbig Be star, MWC297}
\author{KENJI HAMAGUCHI, HIROSHI TERADA, AYA BAMBA, AND KATSUJI KOYAMA\altaffilmark{1}}
\affil{Department of Physics, Faculty of Science, Kyoto University, 
Kitashirakawa-oiwakecho, Sakyo-ku, Kyoto 606-8502;\\
kenji@cr.scphys.kyoto-u.ac.jp, terada@cr.scphys.kyoto-u.ac.jp,  
bamba@cr.scphys.kyoto-u.ac.jp, koyama@cr.scphys.kyoto-u.ac.jp}
\altaffiltext{1}{CREST: Japan Science and Technology Corporation (JST), 4-1-8 Honmachi, Kawaguchi, Saitama 332-0012}

\begin{abstract}
 Hard X-ray emissions from a Herbig Be star 
MWC297 were discovered in three separate observations spanning 5 days 
in April 1994 with the Advanced Satellite 
for Cosmology and Astrophysics (ASCA).
An X-ray flare was found at the beginning of the
second observation with a maximum luminosity of $\approx$ 4.9 $\times$ 10$^{32}$ ergs/s, 
which is five times larger than that of the quiescent phase (the first observation). 
It then declined with an $e$-folding time of $\approx$ 5.6 $\times$ 10$^{4}$ sec to 
the pre-flare level in the third observation.
The X-ray spectra are explained by absorbed thin-thermal 
plasma models.  The temperature in the quiescent phase of $\approx$ 2.7 keV
is significantly higher than that of main-sequence OB stars 
and similar to low mass young stellar objects (YSOs) and other 
Herbig Ae/Be stars observed with ASCA.
 The temperature increased in the flare phase to about 6.7 keV at the flux maximum, 
then decreased to 3.2 keV in the decay phase. 

These facts strongly suggest that X-rays from Herbig Ae/Be stars, at least for
MWC297, are attributable to magnetic activity similar to low mass YSOs.
Since no theory predicts surface convection zone in massive stars like
MWC297, our results may require a mechanism other than the conventional  
stellar dynamo theory.
Possible magnetic activity could be either the stellar interior shear
or the inherited magnetic field from the parent molecular cloud.

\end{abstract}

\section{Introduction}
 	The origin of the X-ray emission from main sequence stars (MSSs) is currently considered to 
be either stellar wind shocks for massive stars (O to B5, e.g. Lucy \& White 1980)
or magnetic activity for low mass stars (F5 to M). 
The former is responsible to a strong stellar wind, whereas the latter is a dynamo
process, which is due to coupled effects of the surface convection and the differential rotation.   
Since intermediate mass stars have neither a strong stellar wind nor a surface convection zone,
they exhibit only faint or no X-rays.

Essentially, the same scenario has been successfully applied to pre-main-sequence stars (PMSs).
Low mass PMSs called T-Tauri Stars (TTSs) and protostars 
have been found to be strong X-ray emitters, possibly due to the enhanced solar-like magnetic activity
(Feigelson \&  Kriss 1981; Montmerle et al. 1983).
Intermediate mass of PMSs are called Herbig Ae/Be stars (HAEBEs) and lack both a strong stellar wind and 
a surface convection layer (e.g. Palla 1999).  Consequently, strong X-ray emission has not been predicted 
from Herbig Ae/Be stars.  

Zinnecker \& Preibisch (1994), however, reported ROSAT X-ray detections from 11 HAEBEs 
out of 21 samples, whose spectral types ranged from B0 to F5.  These facts
force us to suspect a hidden low mass companion with the HAEBEs.   
However, the ROSAT results  show  a  correlation between X-ray luminosity ($L_{X}$) to
bolometric luminosity ($L_{bol}$), and the observed luminosity of $10^{30}-10^{32}$  erg/s 
is significantly larger than those of typical low mass PMSs of $10^{28}-10^{30}$ erg/s.  
Thus, the hypothesis of a hidden low mass companion is unlikely.
Zinnecker \& Preibisch (1994) also found no correlation of $L_{X}$ to the stellar rotation velocity ($v$ sin$i$), 
but found a weak correlation with the velocity of stellar wind ($v_{\infty}$) and the mass loss rate 
($\dot{M}$).  These facts  favor the stellar wind origin for the HAEBE X-rays.
However, the ratio of X-rays to bolometric luminosity $(L_{X} /L_{bol})$ is in the range of 
$10^{-4}-10^{-6}$, larger than those of massive MSSs by one to three orders.   

The puzzle of X-ray emissions from HAEBEs  may be related to
the "paradox"  in the IR band.   Bohm \& Catala (1994) found many similarities 
between TTSs and HAEBEs and  speculated that very young HAEBEs still surrounded by 
circumstellar accretion disks and embedded deeply in the cloud may have jets or outflows,
possibly due to magnetic activity.

Spectral and timing studies with wide energy bands are essential
to address the origin of the HAEBE X-rays, hence, ASCA observations on some selected HAEBEs 
have been performed. Skinner \& Yamauchi (1996) found a high temperature component above 1.6 keV 
from HD104237 (A4e), whereas Yamauchi et al. (1998) detected a deeply embedded Herbig Ae star IRAS12496-7650 with 
extremely large N$_{H} \sim 2 \times 10^{23}$ Hcm$^{-2}$ and high temperature plasma of $\approx$ 2.7 keV.
This high temperature plasma can not be produced by shocks of 
stellar wind of HAEBEs with a typical velocity of $100-500$ km/s.

For further study, we selected a highly reddened star MWC 297 with extremely strong 
Balmer and silicate lines as one of the best samples.  Hillenbrand et al. (1992) reported that MWC297 
is a pre-main-sequence star with a spectral type of O9, at a distance of 450pc
 and the age of $\sim$ 3 $\times$ 10$^{4}$ years.  Drew et al. (1997), however, predicted it 
to be a B1.5 zero-age main-sequence star at a distance of $\sim$ 250 pc.
They also argue that MWC297 is a rapid rotator with $v$ sin$i \sim$ 350 km/s.
The visual extinction is accepted as $8-10$ (McGregor, Persson, \& Cohen 1984; Hillenbrand et al. 1992;
Porter, Drew, \& Lumsden 1998) though Berrilli et al. (1992) and Hou, Jiang, \& Fu (1997) measured lower A$_{V}$ ($\sim$3).
	
	This paper reports the ASCA observation of the Herbig Be star MWC297, presents the first detection of 
a large X-ray flare, and gives some constrains on the emission mechanisms.
We adopt the distance derived by Hillenbrand et al. (1992), but the conclusion is essentially
the same as the case using the results of Drew et al. (1997).

\section{Observations and Data Reduction}

	ASCA observed MWC297 three times spanning 5 days in April 1994, pointing at the same position of 
($\alpha_{2000}, \delta_{2000}$) = (18h27m51.1s, $-$3$^{\circ}$45$'$28.1$''$).
 The observation log is shown in Table \ref{OBS_LOG}.

	ASCA is equipped with four thin-foil X-ray telescopes (XRT) providing a cusp-shape beam 
with a 3$'$ half power diameter.
Two X-ray CCD cameras (SIS0 and SIS1) and two position sensitive gas counters 
(GIS2 and GIS3) are installed on each of the focal planes.
The SIS and GIS are sensitive in the 0.4$-$10 keV and 0.8$-$10 keV band, respectively.
The data from SIS and GIS were collected in the 1-CCD faint mode and normal PH mode, respectively.

	We used the FTOOLS package version 4.0 for the data reduction. 
The data were screened with the normal criteria to exclude the data during 
the South Atlantic Anomaly, earth occultation and low geomagnetic rigidity.
Particle background is rejected by the pattern matching for the SIS and 
by the pulse rise time selection for the GIS.

\section{Analyses and Results}

\subsection{Image and Temporal Variation}

In Fig. \ref{IMAGE}, we made the X-ray image in the 0.8$-$10 keV band
using all the available data from GIS2 and GIS3 in the three observations (see Table \ref{OBS_LOG}). 
Two point-like sources are found with the criterion of S/N $>$ 5.0 : the brighter source at
($\alpha_{2000}, \delta_{2000}$) = (18h27m40.6s, $-$3$^{\circ}$49$'$50.6$''$)
and the fainter source at (18h27m17.7s, $-$3$^{\circ}$40$'$1.5$''$).  
Within the typical ASCA position error
of $\sim$ 40$''$ (Gotthelf 1996), two objects MWC297 and SH2-62
are found from the SIMBAD database at the position of the brighter source;
SH2-62  is a HII region surrounding the MWC297 (Sharpless 1959; Chen, Gao, \& Xiong 1995).
In the error region of the fainter source, located north west of MWC297, we
find only one object IRAS18245-0342 in the SIMBAD database.

We extract X-ray events from the brighter source from a 3$'$ radius circle. 
Fig. \ref{LIGHTCURVE} is the X-ray light curve in the full energy band, merging the  SIS 
(0.4$-$10 keV) to the GIS (0.8$-$10 keV) data.
The flux in the first observation is found to be constant by the $\chi^{2}-$test, 
but is increased by a factor of five at the beginning of the second observation, 
then gradually decreases  exponentially to nearly the same flux of the pre-flare level
at the end of the third observation.
This short time variability indicates that the X-ray emitting region is compact, 
hence is attributable to Herbig Be star MWC297 rather than an extended object SH2-62.
Here and after, we regard the brighter source to be MWC297, and other remote 
possibilities such as uncatalogued low mass young stellar objects (YSOs), or hidden low mass companions 
are separately discussed in Section \ref{subsec : lowmass}.

 The time variability with an exponential-decay is similar to that of a flare
from low mass stars.  We accordingly regard that we observed a decay phase of an X-ray 
flare during the second and third observations, whereas 
the first observation fell on a quiescent phase. The on-set of the flare is 
between the first and second observations.
  We fit the light curve to a constant plus exponential decay model and
obtained the best-fit $e$-folding time of $\approx$ 5.6 $\times 10^{4}$ sec (Table \ref{CURVE_PARAM}).

\subsection{Spectra}
  The X-ray spectra of MWC297 are obtained
by extracting the X-ray events from the same region as the data of the light curve, whereas 
the background is estimated from an annulus with 3$'$ inner radius
and 6$'$ outer radius centered on MWC297.  The background subtracted 
SIS and GIS spectra for each observation are separately displayed in Fig. \ref{SPECTRA}.
For the spectra of each observation, we have made  model fittings  simultaneously to the SIS and GIS spectra.

\subsubsection{one-temperature plasma}

We tried a thin thermal plasma model (Raymond-Smith model) with interstellar absorption.  Acceptable fits 
are obtained for all the spectra of the three observations. 
The best-fit parameters are listed in Table \ref{BEST_FIT_MWC297}. 
The column density of hydrogen is 
consistent with a constant value during the three observations.
  According to the relation $N_{\rm H}$ = 2.2 $\times$ 10$^{21} A_{V}$ cm$^{-2}$ 
(Ryter 1996; Cardelli, Clayton, \& Mathis 1988),
the mean $N_{\rm H}$ value of $\approx$ 2.4 $\pm$ 0.9 $\times$ 10$^{22}$ cm$^{-2}$ is converted
to the visual extinction of  $\sim$ 10 $\pm$ 4, which is in excellent agreement with 
the observed value of Av = 8$-$10.

The plasma temperature in the quiescent phase of 2.7 keV 
was increased to 6.7 keV at the flare and then decreased to 3.2 keV.
This would be the first detection of a flare for early-type stars, accompanied with a plasma heating 
in the rising phase, and cooling in the decaying phase.
The flare peak, however, is missed; we only detected X-rays
after the on-set of the flare.  Thus, the peak luminosity should be larger than
4.9 $\times$ 10$^{32}$ ergs/s, which is the maximum value found at the beginning of the second observation.
Then, the total energy released in the X-ray band is estimated to be 
more than 2.7 $\times$ 10$^{37}$ ergs (0.4$-$10 keV).

The abundances in the first and third observations are
consistent with one solar, but that in the second observation is significantly lower than one solar.\\

\subsubsection{multi-temperature plasma} 
Since X-ray spectra of solar-mass PMSs and brighter Herbig Ae stars
often exhibit a multi-temperature structure with a cool X-ray component, which 
is naturally interpreted as a quiescent coronal component for the solar mass stars, 
and a hard component which is responsible to a flare 
(e.g. A Herbig Ae star, HD 104237 in Skinner \& Yamauchi 1996; A classical TTS, 
SU Aur in Skinner \& Walter 1998 ), 
we also tried a two-temperatures model (two Raymond-Smith model) with common absorption,
although a simple one-temperature model is already acceptable with the $\chi^{2}$ test.

At first, we allowed the two temperatures and common N$_{H}$ to be free parameters 
for each abundance set of the hotter component and cooler component.
For the abundances of hotter and cooler components, we tried for three cases: (1) fixed to those of the best-fit 
one-temperature plasma results, and those of solar, (2) free parameters and fixed to one solar, and 
(3) both free parameters.
However, we find no significant improvement of $\chi^{2}$ values from a one-temperature model
as given in Table \ref{2T_FIT_MWC297}.
Also, the cool component is found to contribute only a small fraction even in the 
soft X-ray band (0.4 $-$ 2 keV), or conversely, the best-fit temperatures of the 
hotter component are essentially the same as those found in the one-temperature fittings.

We should note that for only the flare spectrum in Case (3), the soft component contributes a significant fraction.
However, the temperatures of hotter and cooler components are about 1 keV and 17 keV, respectively, 
and abundances for a cool component are found to be zero. 
These results are unrealistic, and are out side of an allowable range 
to be accepted as a two-component plasma concept for stellar plasmas. 

Next, we fixed the N$_{H}$ value to that expected from the optical extinction 
(N$_{H} = $ 1.83 $\times$ 10$^{22}$ cm$^{-2}$) 
and proceeded with two-temperature fittings.  In this case, the data (spectra) require no soft X-ray 
component at all; the best-fit normalizations of the soft components are zero. 

No improvement for the two-temperature model would be due to the limited photon counts;
in the flare, SIS0+SIS1 and GIS2+GIS3 counts are only  $\sim$1200 and  $\sim$800 photons, respectively, 
whereas those in a quiescent state (first observation, for example) are $\sim$400 (SIS0+SIS1) 
and $\sim$200 (GIS2+GIS3) photons.
Also, the large column density of hydrogen to the source reduces most of the soft X-ray photons, hence the observed 
spectra become insensitive to a cool component, even if it exists.  Thus we conclude that the present 
data set have no capability to judge whether it includes a cool component or not.  
Accordingly, we continue our discussion on the hot component, which parameters are essentially the same 
from both the one- and two-temperature model fittings.

\subsection{IRAS18245-0342}
 Since X-ray emission from an unclassified IR source, IRAS18245-0342 
(Beichman et al. 1988; David et al. 1993) is serendipitously discovered 
in the GIS images, we briefly report the X-ray properties. 
The data are obtained from a 3$'$ radius circle, and the background 
is taken from a source free region.
X-rays are found in the second and third observations, and the spectra are 
separately shown in Fig. \ref{SPECTRA_S2}.  
Since the flare epoch of IRAS18245-0342 coincides with that of MWC297, one may argue that 
possible contamination from MWC297 makes an artificial flare of IRAS18245-0342. We thus made an 
X-ray light curve of a blank sky at the same angular distance from MWC297, then found 
no contamination, nor any flare-like event from the blank sky.  
Therefore, we conclude that the flare is not an artifact but is really coming from the IRAS18245-0342. 
We then fit the spectra to a bremsstrahlung model with interstellar
absorption. The best-fit parameters are shown in Table \ref{BEST_FIT_S2}.
We see consistent results for the column density between the second and third observations 
although the errors are large.  We then fixed to an averaged value (1.5 $\times$ 10$^{22}$ cm$^{-2}$)
and obtained acceptable fittings again with the parameters listed in Table \ref{BEST_FIT_S2}.

Although 90\% errors are still overlapped, the temperature is lower in the third observation. 
Thus, we suspect that IRAS18245-0342 also shows a flare with a cooling process.  

Assuming that IRAS18245-0342 is the same member of the star forming region with MWC 297, hence at the same distance, 
the flare luminosity is estimated to be 6.5 $\times$ 10$^{31}$ ergs/s (0.4$-$10keV).  A flare
with this luminosity, together with the moderate absorption of N$_{H} \sim$ 10$^{22}$ cm$^{-2}$, is consistent with the 
scenario that IRAS18245-0342 is an embedded low mass YSO.

\section{Discussion}

  We found a hard X-ray flare from the position of MWC297. The flare phenomena of 
the first rise associated with an increase of temperature, and slow exponential decay
associated with an apparent cooling resemble to those from low mass YSOs.  We thus discuss
whether the X-rays are due to the Herbig Be star MWC 297, and what is the emission mechanism and other relevant issues.

\subsection{Possibility of X-rays from Low Mass Stars}
\label{subsec : lowmass}
  Testi, Palla, \& Natta (1998) found  about twenty sources in the ASCA error circle around  MWC297 
with the K-band absolute magnitude M$_{K}$ brighter than 5.2. 
The star formation activity probably occurs simultaneously in a dense cloud core, hence 
the ages of these nearby stars would be the same as MWC297.
The age of MWC297 is estimated to be $\sim$ 10$^{5}$ years, which is equivalent 
to the protostar phase in low mass stars.
Thus, most of the IR sources are likely to be low mass protostars.
Low mass protostars, at least some fractions, are known to exhibit highly absorbed  
hard X-rays even in the quiescent phase 
(Koyama et al. 1996; Kamata et al. 1997; Tsuboi et al. 1999), which
resemble those found from MWC297.
Are present X-rays due to a protostar?
The luminosity of a protostar in the quiescent phase is typically less than 
10$^{31}$ ergs/s, which is less than 10\% of the present MWC297 luminosity.

Integrated flux from many protostars may  account for the observed high flux, 
but a simple time profile of the flare can not be reproduced by integrated 
emission from many objects.  The flare peak and total flare energy 
of the present observations are more than 4.9 $\times$ 10$^{32}$ ergs/s and 2.7 $\times$ 10$^{37}$ ergs, respectively.
Both values largely exceed those of the flare from protostars. 

Thus, quiescent and flare X-rays are not attributable to protostars, regardless of single or 
integrated emission of many protostars.

  We found that the X-ray determined N$_{H}$ to MWC297 is in good agreement 
with the visual extinction of MWC297.  
The millimeter continuum map taken by Henning et al. (1998) which traces dense 
and cold envelopes 
has a clear peak at the position of MWC297, hence the N$_{H}$ we observed is 
locally concentrated near to MWC297.
Furthermore, no companion star with MWC297 is found from the binary search studied by Pirzkal, Spillar, \& Dyck (1997)
although the probability to detect an existing companion is not high ($\sim$ 23\%).
These facts lead us to conclude that X-rays are really attributable to the Herbig Be star MWC297.

\subsection{Emission Mechanism}

 The luminosity ratio of X-rays to bolometric (L$_{X}$/L$_{bol}$) of massive stars
is typically 10$^{-6}-$10$^{-8}$ (Long \& White 1980; Chlebowski, Harnden, \& Sciortino 1989).
The Lx/L$_{bol}$ ratio of MWC297 in the quiescent phase is 3 $\times$ 10$^{-7}$, which is 
a typical value for massive stars.
 
 This may support that X-rays from MWC297 are due to the same mechanisms of those of 
massive stars; the stellar wind shock model.  
Nisini et al. (1995) derive the terminal velocity of MWC297 as 380 km/s 
from the flux ratio between the 3.6 cm radio continuum and the Br$\alpha$ line emission.
Then, the plasma temperature can be estimated to be at most $\sim$ 0.25 keV, which is
almost one order of magnitude smaller than those of the observed value of $\approx$ 2.7 keV in the 
quiescent phase.
We note that the stronger stellar wind with higher velocity in more massive stars produce 
a plasma temperature typically less than 1 keV (Seward et al. 1979; Harnden et al. 1979).

  Also, the rapid X-ray variability in a day detected in MWC297 is unusual
for massive stars with an exception of an O9.5 star $\zeta$-Orionis, which showed 
a small increase of the X-ray flux by  $\sim$ 30\% in 2 days (Berghofer \& Schmitt 1994a, 1994b). 
This can be explained by the stellar wind model that 
fast-moving material impacts on slowly moving denser shells.
The variation of the X-ray intensity is due to the fluctuation of the emission measure.
However, this scenario predicts no significant change of the plasma temperature during the flare,
hence it is not the case for MWC297.

Thus, the stellar wind model is unlikely for X-ray emission from MWC297 in both 
quiescent and flare phases.
We note that high temperature plasma with $\sim$ 2 keV in HAEBEs is also reported 
in HD104237 (Skinner \& Yamauchi 1996) and IRAS 12496-7650 (Yamauchi et al. 1998). 

Infrared observations suggest several similarities between HAEBEs and low mass PMSs.
X-ray properties of MWC297 and other HAEBEs also resemble those  of low mass PMSs.
The plasma temperature of the quiescent phase (1$-$3keV for low mass PMSs) is the same with 
that of MWC297 and other HAEBEs ($\sim$ 2 keV).
  Low mass PMSs occasionally exhibit  rapid flares with the shape of fast-rise and 
exponential-decay, whose peak intensity is  about 2$-$20 times as large as in the quiescent phase 
(e.g. Montmerle et al. 1983; Koyama et al. 1996; Tsuboi et al. 1998).
 The flare of MWC297 also has a strong peak at least five times as large as the quiescent phase
followed by an exponential-decay, although the $e$-folding time of MWC297 ($\sim$ 15 hrs) is
longer than those of low mass PMSs of a few hours. 
 We note that an X-ray flare found from a classical Be star, $\lambda$-Eridani 
(Smith et al. 1993) also shows a long duration of about 50 ksec.

 These indicate that X-rays from MWC297 are, like those of low mass PMSs,
due to magnetic activities.  
  Assuming radiative cooling for the flare event, we evaluate the 
electron density (n$_e$) and the volume of the plasma (V) from the equations of
Lx = n$_e^{2}\Lambda(T)V$ and $\tau_{cool}$ = 3n$_{e}kT/n_e^{2}\Lambda(T)$ 
($\Lambda(T)$, the emissivity;~$\tau_{cool}$, the cooling time scale of the plasma).
The electron density (4.4 $\times$ 10$^{10}$ cm$^{-3}$) is typical for solar and low-mass stellar flares.
The plasma volume is 1.8 $\times$ 10$^{34}$ cm$^{3}$, and a typical length (V$^{1/3}$) is 2.6 $\times$ 10$^{11}$ cm. 
It is smaller than the radius of MWC297 (6.3 $\times$ 10$^{11}$ cm, Hillenbrand et al. 1992), 
and the magnetic plasma loop, for example, can be located on the stellar surface.

The problem of the magnetic activity scenario, however, is that massive stars do not have a surface 
convection zone, which is required for a conventional dynamo process (e.g. Palla 1999).
  An alternative mechanism for a magnetic amplification is 
that the disk shear instability and the turbulence in the accretion disk also act as dynamo 
(Brandenburg et al. 1995).
The infrared and radio observations of Herbig Be stars generally 
show no evidence of a disk (Natta, Grinin, \& Mannings 1999).
Neither outflow nor an optical jet are seen around MWC297 itself.
Hence, this scenario is controversial for the case of MWC297.

``Fossil'' magnetic field inherited from the parent molecular 
cloud (Tayler 1987) or shear dynamo activity by 
tapping the initial stellar differential rotation (Tout \& Pringle 1995) is more likely.

\subsection{Abundance Variation}

The flare plasma of MWC297 shows lower abundance than that in the quiescent phase although their
90\% confidence errors overlap with each other.
  Similar abundance variations have been reported in other stars 
(V773 tau in Tsuboi et al. 1998; Algol in Ottmann \& Schmitt 1996; UX-Ari in Tsuru et al. 1989; 
the sun in Sylwester, Lemen, \& Mewe 1984).  
However, no conclusive scenario for the apparent abundance variations has been proposed.

  The flare spectrum is also fitted to a power-law model (See Table \ref{BEST_FIT_MWC297}). 
Hence, a possibility of non-thermal X-ray emission from  stellar plasma can not be excluded.
 Non-thermal bremsstrahlung  X-rays above 20 keV are found with 
the Japanese solar X-ray observatory YOHKOH, which are explained 
by the collision of accelerated electrons in the flare loop to the solar surface (Sakao, Kosugi, \& Masuda 1998).
If a significant fraction of the non-thermal X-rays are produced in a flare, the spectrum 
would show apparent low abundance.

\section{Conclusions}

We presented and analyzed the ASCA data in the quiescent and flare phases of MWC297, 
and derived the physical condition of the X-ray emitting plasma.
The results are as follows:

\begin{itemize}
\item We detected two X-ray sources with S/N $>$ 5.0, whose 
 coordinates are in agreement with those of MWC297 and IRAS18245-0342, respectively.
\item The light curve of MWC297 showed a powerful flare with the peak flux at least five times 
as large as that 
of the quiescent phase, followed by a gradual decay of the $e$-folding time of $\approx$ 5.6 $\times 10^{4}$ sec.
\item All spectra are well explained by absorbed thin-thermal plasma models.
 The hydrogen column density is almost constant at $\approx$ 2.4 $\times$ 10$^{22}$ cm$^{-2}$, 
  which is entirely consistent with the visual extinction of MWC297.
 The luminosity in the quiescent phase ($\sim$ 10$^{32}$ ergs/s) is larger than those of 
low-mass stars.
 X-rays from unresolved low mass companions or nearby stars are unlikely. 
\item The plasma temperature in the quiescent phase is 2.7 keV, which is similar to other HAEBEs. 
It is heated up to 6.7 keV in the flare and then cooled down to 3.2 keV.
The stellar wind model, which is common in massive stars, can not explain these 
phenomena found in MWC297.
\item The coronal X-ray emission by magnetic activity is likely for the X-ray emission mechanism of 
MWC297.   The typical length of the flaring plasma (2.6 $\times$ 10$^{11}$ cm) is smaller than the radius 
of MWC297.
The magnetic amplification mechanism, however, is puzzling.
Magnetic activity on the stellar surface by  the stellar internal shear or ``fossil'' magnetic field 
inherited from the parent cloud are possible scenarios.

\item The abundance during the flare is lower than the solar and those in the quiescent phase.
      
\item IRAS18245-0342 also shows a flare-like time variation, suggesting a new low mass YSO.
\end{itemize}

  Three separate observations spanning about 5 days enable us to obtain both the flare and quiescent data, 
but the light curve to cover full phases of the flare was not obtained.
Long-time continuous observations with the next generation of satellites, Astro-E, XMM and Chandra, which have 
a large effective area and/or high energy resolution will reveal the true emission mechanism of Herbig Ae/Be stars.

\acknowledgments
We thank all members of the ASCA team and administrators of 
the SIMBAD astronomical database operated by CDS in Strasbourg, France.
S. Yamauchi, T. Sakao and an anonymous referee are greatly appreciated for their useful comments and critiques.

\clearpage

\figcaption[f1.eps]{The GIS2 + GIS3 image in 0.8$-$10 keV band.
The image are accumulated from the three observations and are smoothed by the 
GIS point spread function.
The contour levels are a logarithmic scale; the counts of GIS2 + GIS3 per pixel
are 1.0, 1.7, 3.0, 5.5, and 10.1, respectively.
The outer circle is the GIS field of view (FOV) and the inner square is SIS0 
chip1 (SIS1 chip3) FOV.  IRAS18245-0342 is out of view for SIS. \label{IMAGE}}

\figcaption[f2.ps]{The light curve of MWC297. 
The vertical axis is the averaged count rate of four detectors (SIS 0,1 ; GIS 2,3).
The horizontal axis is the time after the beginning of the first observation.
1 bin is 2048 seconds. \label{LIGHTCURVE}}

\figcaption[f3.eps]{SIS (left panels) and GIS (right panels) spectra of MWC297.
Upper panels are those of the 1st observation, middle panels are 2nd and lower panels are 3rd. 
The best-fit Raymond-Smith models of simultaneous fittings are shown by the solid line, 
and residuals from the best-fit values are displayed in the lower section of each panel.
\label{SPECTRA}}

\figcaption[f4.eps]
{GIS spectra of IRAS18245-0342. The left panel is those of the 2nd observations, and right panel is 3rd.
The best-fit Raymond-Smith models of simultaneous fittings are shown by the solid line, 
and residuals from the best-fit values are displayed in the lower section of each panel.
\label{SPECTRA_S2}}

\clearpage

\begin{deluxetable}{cllllr}
\tablenum{1}
\tablewidth{0pt}
\tablecaption{ASCA Observation log \label{OBS_LOG}}
\tablehead{
\colhead{Observations}   & \multicolumn{2}{c}{Start} & 
\multicolumn{2}{c}{End} &
\colhead{Exposure} \\
   	 		 & \colhead{Date}&
\colhead{Time}	 	 & \colhead{Date}&
\colhead{Time}           & \colhead{ksec}
}
\startdata
1st   & 94/4/~8 & 00:02 & 94/4/~8 & 11:11 & 10.3~~~        \\
2nd   & 94/4/12 & 03:31 & 94/4/12 & 09:26 &  5.0~~~        \\
3rd   & 94/4/12 & 23:53 & 94/4/13 & 12:31 & 12.3~~~        \\
\enddata
\end{deluxetable}

\begin{deluxetable}{lllllr}
\tablenum{2}
\tablewidth{0pt}
\tablecaption{Fitting parameters of the light curve \label{CURVE_PARAM}}
\tablehead{
   			&			& 
Quiescent phase		& \multicolumn{2}{c}{Flare phase} \\
\colhead{Observations}   & 			&
\colhead{1st}	 	& \colhead{2nd \& 3rd}  &
\colhead{2nd}
}
\startdata
Constant        & [$\times$10$^{-2}$~cts~s$^{-1}$]~~~&~~2.27 	     &~~0.22 &~~\nodata\\ 
 		& 			      &~~(2.14--2.39)  &~~($<$0.99) &~~\nodata\\ 
$e$-folding time& [$\times$10$^{4}$~sec]~~~&~~\nodata     &~~5.56 &~~4.59 \\ 
		&			   &~~\nodata     &~~(4.52--6.97)~ &~~(3.53--6.50) \\ \tableline
$\chi^{2}$/d.o.f&&\multicolumn{1}{c}{5.5/18}&\multicolumn{1}{c}{39.6/29}&\multicolumn{1}{c}{15.3/9}\\
\enddata
\tablenotetext{}{The errors listed in parentheses are quoted for 90\% confidence.}
\end{deluxetable}

\begin{deluxetable}{llllll}
\tablenum{3}
\tablewidth{0pt}
\tablecaption{Spectral Parameters of MWC297 \label{BEST_FIT_MWC297}}
\tablehead{
Observations&			& 
1st		&\multicolumn{2}{l}{2nd}&
3rd\\
Status&   		&
Quiescence&\multicolumn{2}{l}{Flare}&
Flare decay}

\startdata
Absorption    &$N_{\rm H}$~[10$^{22}$ cm$^{-2}$]& 2.6       & 2.5       & 2.9        & 2.0         \\
              &                                	&(2.0$-$3.3)~~~~~&(2.1$-$2.8)&(2.5$-$3.4)~~~~~&(1.5$-$2.5)    \\
Raymond Smith &$kT$~[keV]                  	& 2.7       & 6.7       & \nodata    & 3.2         \\
              &                            	&(2.1$-$4.1)&(4.8$-$9.6)&            &(2.3$-$4.7)    \\
              &abundance~[solar]~~      	& 1.3       & 0.18      & \nodata    & 0.57        \\
              &                            	&(0.26$-$4.5)&($<$0.38) &            &(10$^{-3}-$1.6)\\
Power law     &photon index		   	& \nodata   & \nodata	& 2.0        & \nodata        \\
	      &				   	&           &           &(1.8$-$2.3) &             \\
Lx (0.4$-$10keV)\tablenotemark{\dag}&[10$^{32}$ ergs~s$^{-1}$]& 0.82& 4.5      & \nodata    & 0.75        \\
			&			&(0.79$-$0.84)&(4.4$-$4.6)&		&(0.72$-$0.77)\\
E.M       &[10$^{55}$cm$^{-3}$]            	& 1.1       & 3.5       & \nodata    & 0.88         \\ \tableline
$\chi^2$/d.o.f.&                           	& 37.1/42   &108.5/104  &107.0/105   & 55.1/59     \\ 
\enddata
\tablenotetext{}{The errors listed in parentheses are quoted for 90\% confidence.}
\tablenotetext{\dag}{N$_{H}$-corrected X-ray luminosity at 450pc derived from the best-fit parameter.}
\end{deluxetable}

\begin{deluxetable}{lcclll}
\tablenum{4}
\tablewidth{0pt}
\tablecaption{Reduced $\chi^{2}$ values of spectral models \label{2T_FIT_MWC297}}
\tablehead{
\colhead{Models}	&\multicolumn{2}{c}{Abundance}	&\multicolumn{3}{c}{Observations}\\
			&\colhead{Hot comp.}	&\colhead{Cool comp.}	&
\colhead{1st}		&\colhead{2nd}		&\colhead{3rd}
}
\startdata
1T Raymond Smith&	free&		    \nodata    	   	&	0.8820&		1.044&		0.9335\\
2T Raymond Smith&	1T value fixed&     one solar fixed&	0.8811&		1.046&		0.9266\\
2T Raymond Smith&	free&		    one solar fixed&	0.9003&		1.057&		0.9428\\
2T Raymond Smith&	free&		    free	        &	0.9172&		1.005$^{\dag}$&	0.9596\\
\enddata
\tablenotetext{\dag}{Only in this case, the soft component dominates the hard component and the abundance 
of the soft component goes to zero.}
\end{deluxetable}

\begin{deluxetable}{llllll}
\tablenum{5}
\tablewidth{0pt}
\tablecaption{Spectral Parameters of IRAS18245-0342 \label{BEST_FIT_S2}}
\tablehead{
Observations&			& 
\multicolumn{2}{l}{2nd}		& \multicolumn{2}{l}{3rd}}
\startdata
Thermal Brems.& $kT$~[keV]                   & 2.8       & 3.2          & 1.6      & 1.4      \\
              &                             &(1.1$-$9.7)  &(1.9$-$6.1)~~~~~~     &(0.7$-$4.2) &(0.8$-$2.4)~~~~~~ \\
Absorption&$N_{\rm H}$~[10$^{22}$ cm$^{-2}$]~~ & 1.8       & 1.5 (Fixed)    & 1.3      & 1.5~(Fixed) \\
          &                                 &(0.4$-$4.9)  & \nodata         &(0.4$-$2.6) & \nodata     \\
Fx~(0.4$-$10keV)\tablenotemark{\dag}&[10$^{-12}$ergs~cm$^{-2}$~s$^{-1}$]&2.7 &           & 1.0     &          \\ 
	  &				    &(2.4-3.0)  &	       &(0.94-1.2)	 &  \\ \tableline
$\chi^2$/d.o.f.&                            & 6.5/5     & 6.6/6        & 6.0/5    & 6.1/6  \\ 
\enddata
\tablenotetext{}{The errors listed in parentheses are quoted for 90\% confidence.}
\tablenotetext{\dag}{N$_{H}$-corrected X-ray flux derived from the best-fit parameter.}
\end{deluxetable}

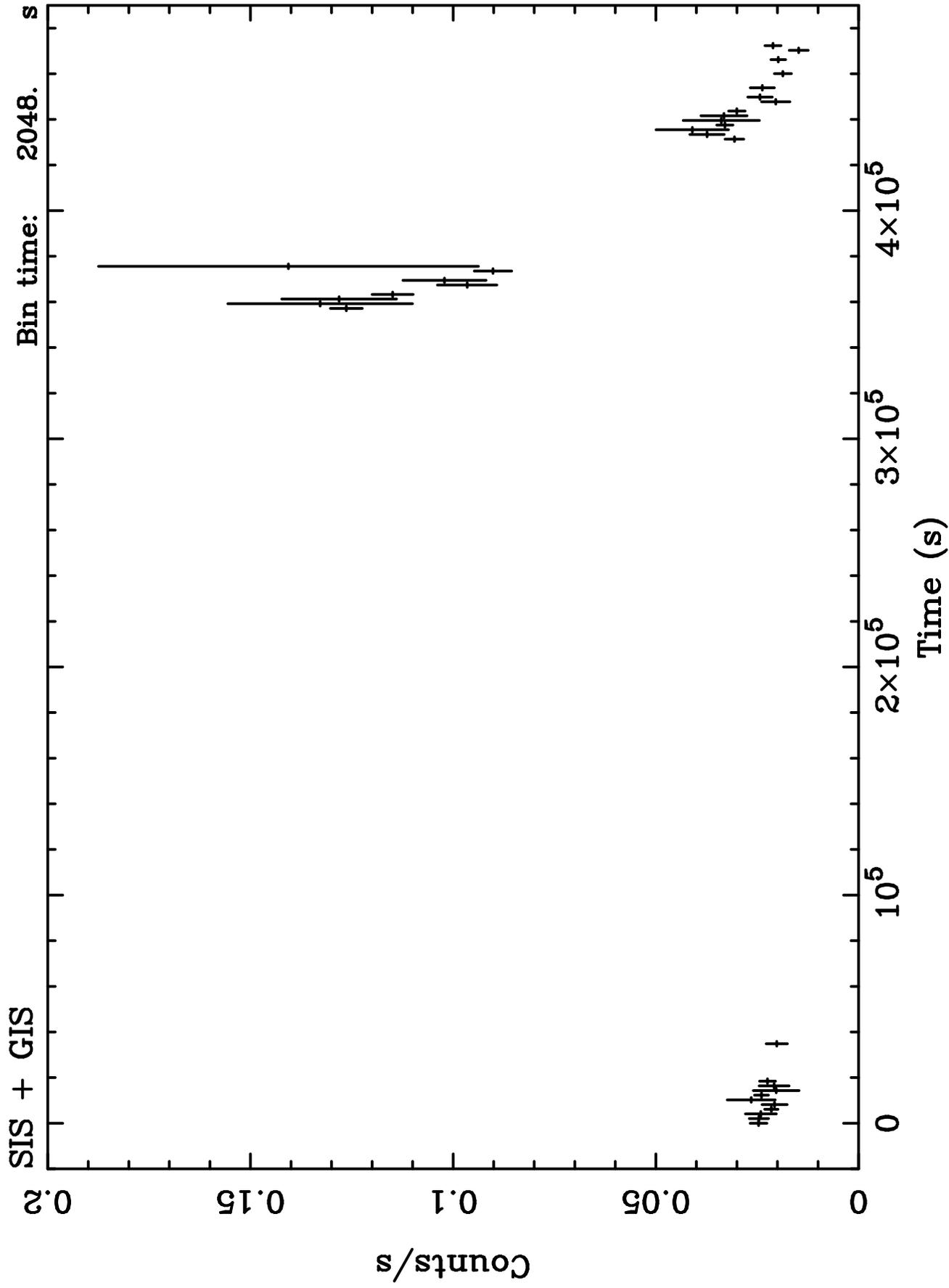
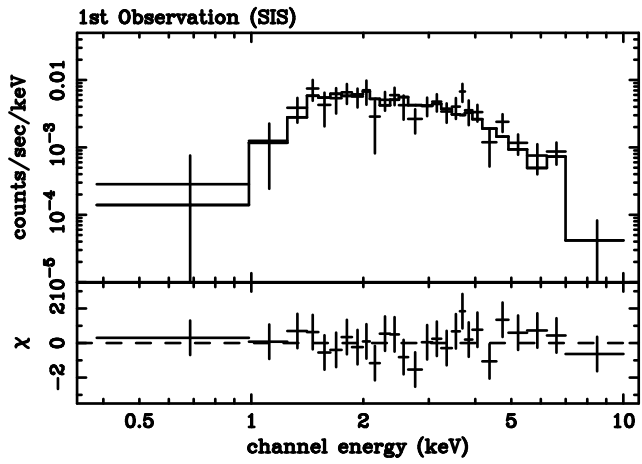
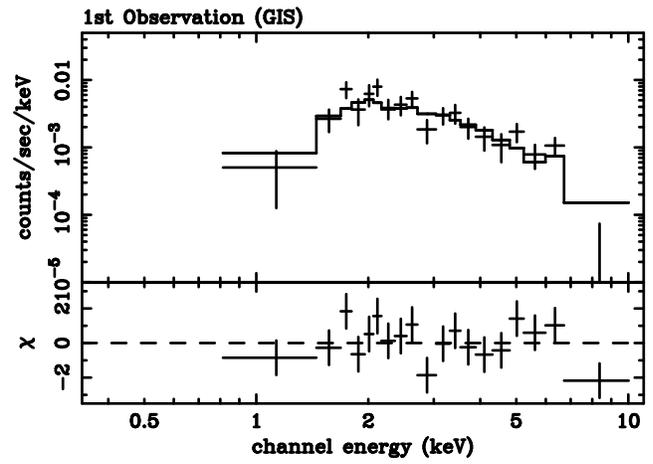
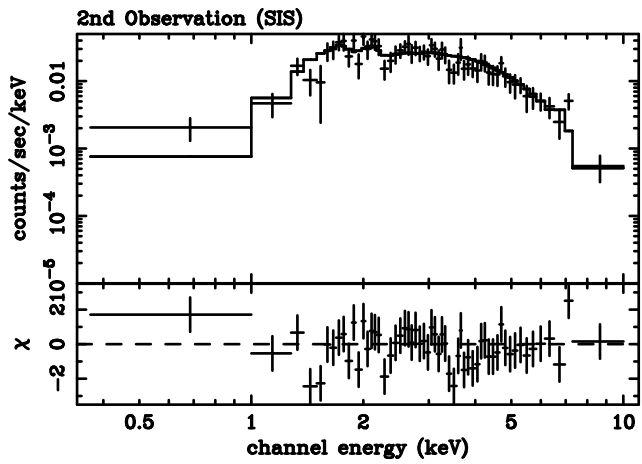
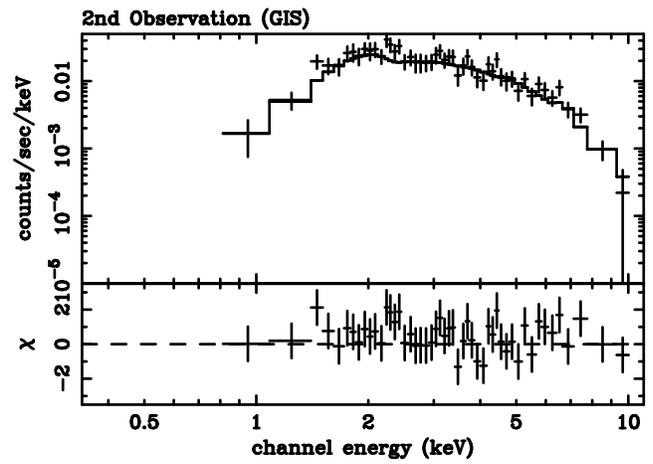
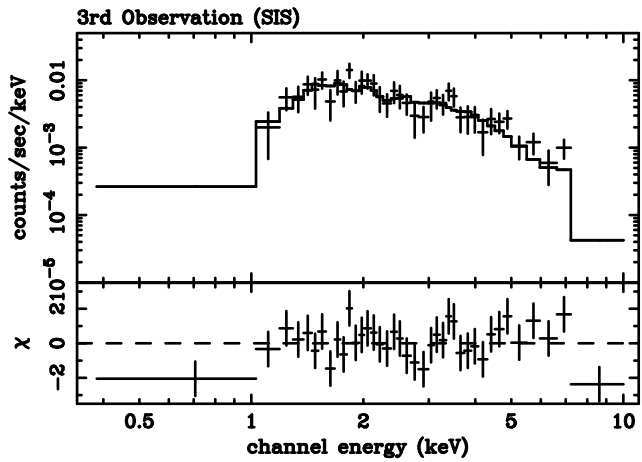
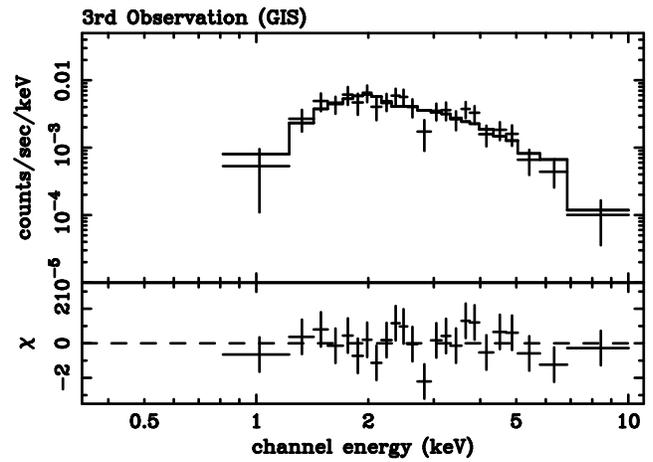
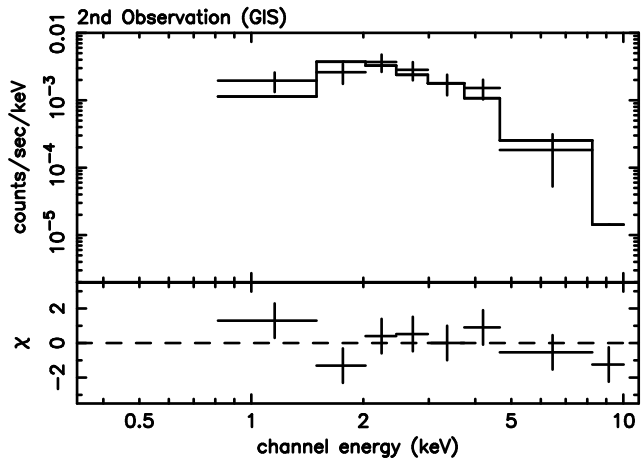
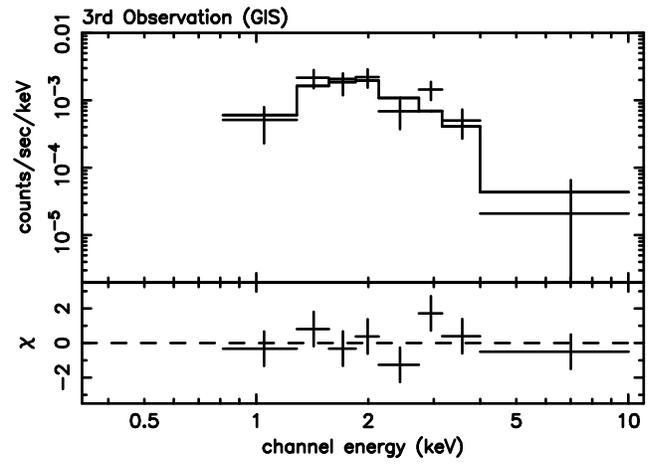
\end{document}